\documentclass[aps,pra,reprint,superscriptaddress,floatfix]{revtex4-2}

\usepackage{graphicx,color}% Include figure files
\usepackage{dcolumn}% Align table columns on decimal point
\usepackage{bm}% bold math
\usepackage{physics}
\usepackage{lipsum}
\usepackage{graphicx}
\begin{document}

\preprint{APS/123-QED}

\title{Ultracold dipolar bosons trapped in atomtronic circuits}

\author{Marc Rovirola}
\affiliation{Departament de F\'isica de la Mat\`eria Condensada,Facultat de F\'{\i}sica, Universitat de Barcelona, 08028
Barcelona, Spain}
\author{Héctor Briongos-Merino}
\affiliation{Departament de F\'isica Qu\`antica i Astrof\'isica,
Facultat de F\'{\i}sica, Universitat de Barcelona, 08028
Barcelona, Spain}
\affiliation{Institut de Ci\`encies del Cosmos de la Universitat de
Barcelona, ICCUB, 08028 Barcelona, Spain}
\author{Bruno Juli\'a-D\'iaz}
\affiliation{Departament de F\'isica Qu\`antica i Astrof\'isica,
Facultat de F\'{\i}sica, Universitat de Barcelona, 08028
Barcelona, Spain}
\affiliation{Institut de Ci\`encies del Cosmos de la Universitat de
Barcelona, ICCUB, 08028 Barcelona, Spain}
\author{Montserrat Guilleumas}
\affiliation{Departament de F\'isica Qu\`antica i Astrof\'isica, Facultat
de F\'{\i}sica, Universitat de Barcelona, 08028 Barcelona, Spain}
\affiliation{Institut de Ci\`encies del Cosmos de la Universitat de
Barcelona, ICCUB, 08028 Barcelona, Spain}

\date{\today}% It is always \today, today,

\begin{abstract}
We consider a ring-shaped triple-well potential with few polar bosons with in-plane dipole orientation.
By diagonalizing the extended Bose-Hubbard Hamiltonian,  
we investigate the ground state properties of the system as we rotate the dipole angle and vary the on-site and dipole-dipole interaction strengths. We find that the anisotropic character of the dipolar interactions, as well as the competition between dipole and on-site interactions lead to different ground states and that the entanglement between sites also depends on the number of particles. We further characterize the system by studying the condensed fraction and coherence properties for different polarization angles, highlighting the possible effect of the dipolar interaction as manipulation tool.
\end{abstract}

\maketitle

\section{Introduction}

The wide set characteristic features of quantum gases, 
such as, entanglement
\cite{FadelScience2018}, transistor-like properties \cite{SingleAtomTransistor,triplewelltransistor} and persistent currents \cite{Viscondi_2011}
can be 
taken advantage
for the development of quantum technologies. 
The atomtronics field studies these quantum phenomena and capitalizes on them to develop quantum devices \cite{RoadmapAtomtronics,Amico2022}. 
To create atomtronic circuits, 
atoms are usually loaded into potential wells or optical lattices that can be designed with many different shapes and potential strengths (links)
that will heavily influence the properties of the system \cite{TunningDipoleGases}.
Triple-well potentials have gained interest due to their simplicity but rich phenomenology \cite{Kumar_Haldar_2019, Wittmann_W__2022, triplewell2, richaud_mixing-demixing_2019}. Two types of set-up are attainable: fully connected wells in a triangular shape or one-dimensional aligned wells. The former being the smallest system to include angular momentum with a superfluid phase that is manifested with vortex currents \cite{Arwas2014,Gallem__2015}.

Ultracold dipolar atoms, such as 
dysprosium~\cite{Klaus2022,Lu2011,Tang2015},
erbium~\cite{BECErbium}, chromium~\cite{Griesmaier2005},
or europium~\cite{Miyazawa2022}, are characterized by a strong dipole-dipole interaction. 
Recently, the realization of a Bose-Einstein condensate of dipolar molecules has also been reported~\cite{Bigagli2023}.
The dipolar interaction exhibits long range and anisotropic behavior leading to new phase transitions \cite{Lahaye_2009, gallemi2013} and fragmentation \cite{Lu2012} in 
atomtronic devices.
Recent experiments using dysprosium atoms have reported, for the first time, the occurrence of vortex states in dipolar Bose-Einstein condensates
by rotating the polarization direction~\cite{Klaus2022}. 
The anisotropic character of the dipole-dipole interaction has been exploited to induce rotation in the system forming quantized vortices, which is a clear signature of superfluidic behavior in a quantum system~\cite{Donnelly}.

Previous studies with polar atoms in triple-well potentials have addressed different quantum phenomena for fixed polarization directions, such as
Josephson-like dynamics \cite{TripleWellJosephson} and  interaction-induced coherence~\cite{Fisher2010}, for one-dimensional aligned wells, and self-trapping \cite{PhysRevA.87.053620} in ring-shaped potentials.
Moreover, protocols for controlling entanglement in similar circuits loaded with dipolar bosons have been proposed ~\cite{Wittmann2023,grun_protocol_2022}.

In this work, we consider few dipolar bosons trapped in an equilateral triple-well potential in a two-dimensional setting. We investigate the static properties of the system by varying the polarization configuration along the in-plane directions and for different on-site and dipole-dipole interaction strengths, extending  a previous analysis done in three specific orientations \cite{gallemi2013}.
We study how commensurable and non-commensurable filling affects the system, and examine how contact interaction drives the correlation between sites, average occupation, condensed fraction and entanglement spectrum.
Beside the discussion of possible phases, here, we analyze the Schmidt gap as a function of the
polarization angle. We show that in the fractional filling case there are angles
for which, independently from the local interaction, there is large
entanglement between the subsystems.

This work is organized as follows: Section \ref{section:model} introduces the system and the extended Bose-Hubbard Hamiltonian. Section \ref{section:properties}
presents the results obtained by diagonalization of the Hamiltonian for different sets of parameters. We discuss the average occupation of the sites for
the ground state for different values of the interaction strengths as a function of the dipole orientation. We discuss the entanglement properties of the ground state in Section \ref{sec-entanglement}. 
In Section \ref{sect-energy} we calculate the degeneration of the ground state and the energy gap as a function of the polarization angle for different values of the interactions.
Finally, Section \ref{section:conclusions} presents the conclusions.

\section{Dipolar Bose-Hubbard Hamiltonian}\label{section:model}

\begin{figure}[htb]
    \centering
    \includegraphics[width=246pt]{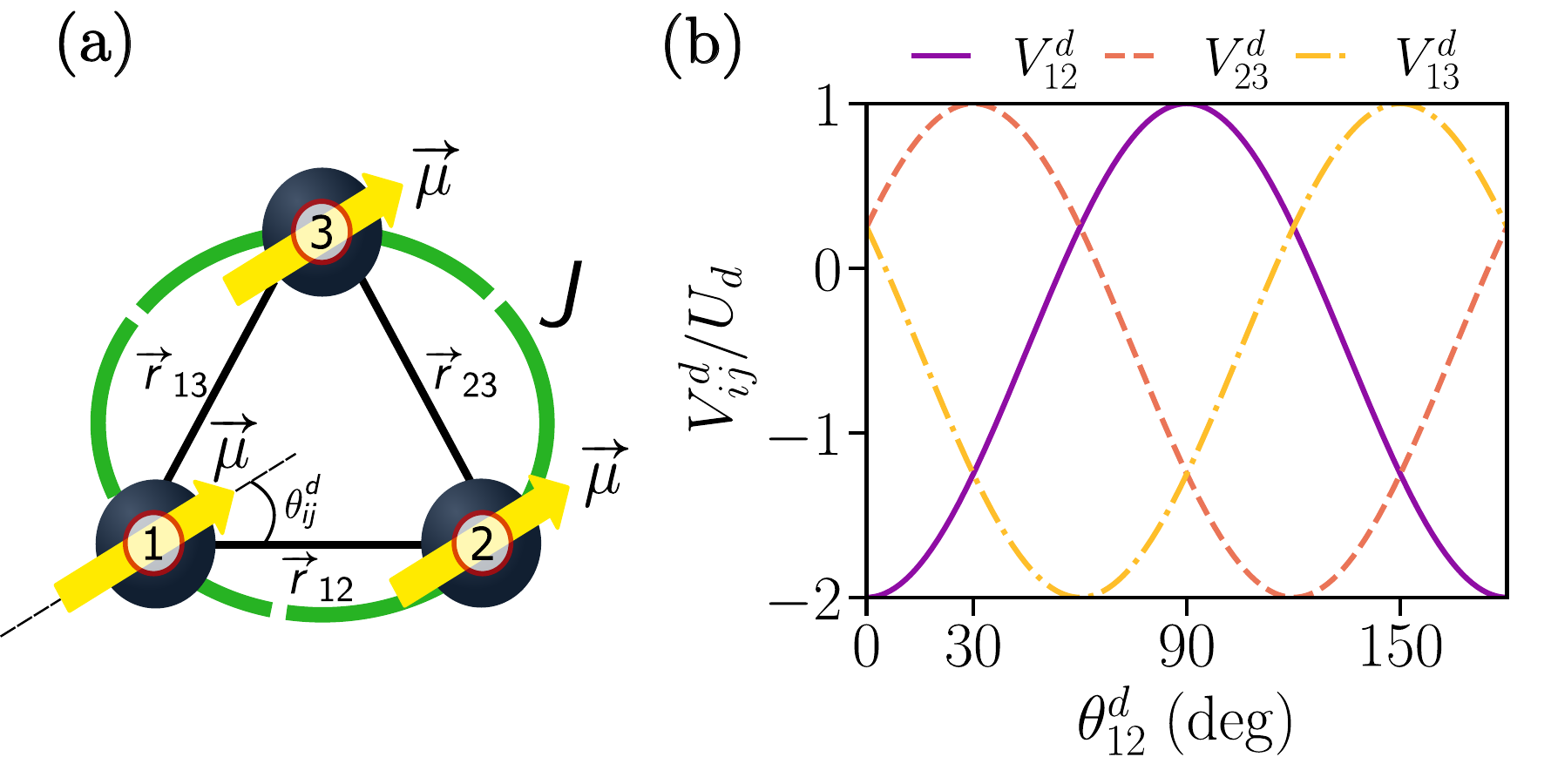}
    \caption{(a) Schematic representation of the equilateral triple well set-up. The yellow arrow represents the effective dipole, $\vec{\mu}$, of each well, the green line represents the tunneling with strength $J$, and $\theta^d_{ij}$ is the angle between the dipole direction and the 
    relative position between sites $i$ and $j$, $\vec{r}_{ij}$. (b) Representation of Eq.~(\ref{equation:Vdd}) as a function of the dipole angle $\theta_{12}^d$, where $\theta^d_{23} =\theta^d_{12} - 60$ deg and $\theta^d_{13} = \theta^d_{12} - 120$ deg.
    }
    \label{tripleWellSetup_Vdd}
\end{figure}

We consider $N$ dipolar bosons confined in a triple-well potential with an equilateral triangular geometry. 
The extended Bose-Hubbard Hamiltonian reads
\begin{equation}\label{hamiltonian}
    {\cal{\hat{H}}} = -J \!\! \sum_{<i,j>}^3 [\hat{a}_i^\dag \hat{a}_j + h.c.] + \frac{U}{2} \!\! \sum_i^3 \hat{n}_i(\hat{n}_i -1) + \sum_{\substack{i,j \\ i \neq j}}^3 V^d_{ij} \, \hat{n}_i \hat{n}_j \,,
\end{equation}
where $\hat{a}_i (\hat{a}_i^\dag)$ are the bosonic annihilation (creation) operators for site $i$, and $\hat{n}_i = \hat{a}_i^\dag\hat{a}_i$ is the particle number operator on the $i$th site.
$J$ is the tunneling rate between neighboring sites, $U$ is the on-site atom-atom interaction that we assume to be repulsive ($U >0$), and $V_{ij}^d$ represents the dipole-dipole interaction strength.
We consider an effective dipole vector $\vec{\mu}$
in each site, whose angle is described with respect to one side of the equilateral triangular shape~\cite{gallemi2013,PhysRevA.102.053306, TripleWellJosephson,Safavi_Naini_2013}. The configuration of the system is schematically shown in Fig.~\ref{tripleWellSetup_Vdd} (a). 

We assume that all the atoms have the same in-plane dipole orientation.
The dipolar interaction between two dipoles located at sites $i$ and $j$ can be written in the following form:
\begin{equation}\label{equation:Vdd}
 V^d_{ij} = U_{d} \, (1-3\cos^2\theta^d_{ij}) \,,
\end{equation}
where $U_{d}$ is the strength of the dipole interaction, independent of $r_{ij}$ due to the symmetry of the system, and
$\theta^d_{ij}$ is the angle between the dipole direction and the relative position between the two dipoles.

Due to the equilateral geometry of the triple-well configuration and the parity of the dipolar interaction, the system is symmetric under 120 deg rotations (and its multiples), preserving its properties, i.e. just moves the labels of the sites. Moreover, due to the symmetric behavior under inversion of the dipolar interaction, studying the local properties of a single site with a dipole direction that ranges from 0 to 180 deg is enough to describe the whole system. From now on, we will consider the in-plane polarization direction with respect to the $\vec{r}_{12}$ direction, defined by the polarization angle $\theta \equiv \theta^d_{12}$, see Fig.~\ref{tripleWellSetup_Vdd} (a) and it will be always expressed in degrees.

The dipolar interaction favors the localization of the atoms in the two sites whose dipoles are more aligned, resulting in an overall lower energy (i.e., their dipolar interaction term is the most negative among the three of them). These favored sites change with the orientation of the dipole; see Fig.~\ref{tripleWellSetup_Vdd} (b). For example, the occupation of the site pair 2-3 is favored for $\theta$ from 90 to 150, making site 1 less favorable to be populated. A special symmetry arises for the angles $\theta = 30, 90$ and  $150$, in which two dipolar interactions have the same strength. In these cases, the pair of favored sites is not well defined, resulting in a non-factorizable ground state in terms of single-site Fock states.

\section{Ground state 
}
\label{section:properties}

The system is characterized by the parameters 
of the Hamiltonian: $J, U, U_d$, as well as $N$ and the in-plane polarization direction.
We obtain the eigenvectors and eigenenergies of the system by exact diagonalization of Eq.~(\ref{hamiltonian}) for a fixed number of atoms and polarization angle $\theta$. 
In our calculations, the tunneling parameter is set to $J/h= 1$~ Hz, and different numbers of atoms $N$ with filling factor $\nu=N/3$ are considered.
The ground state, $\ket{\Psi_{\rm GS}}$, is calculated as a function of the polarization angle for different values of 
the on-site interaction $0 \leq U/J \leq 45$,  
fixing the inter-site interaction strength to $U_d/J=1, 3$ and $10$.

The many-body wave function can be written as a superposition of Fock states as
\begin{equation}
\ket{\Psi_{\rm GS}} = \sum_{n_1,n_2,n_3 = 0}^N C_{n_1,n_2,n_3}\ket{n_1 n_2 n_3} \,,
\end{equation}
where $n_i=0,1,...,N$ is the number of atoms in site $i$, and $|C_{n_1,n_2,n_3}|^2$ is the probability of finding the system $\ket{\Psi_{\rm GS}}$ in the corresponding Fock state $\ket{n_1n_2n_3}$.

\subsection{Average occupation
}\label{density}

\begin{figure}[hb]
    \centering
    \includegraphics
    [width=241pt]{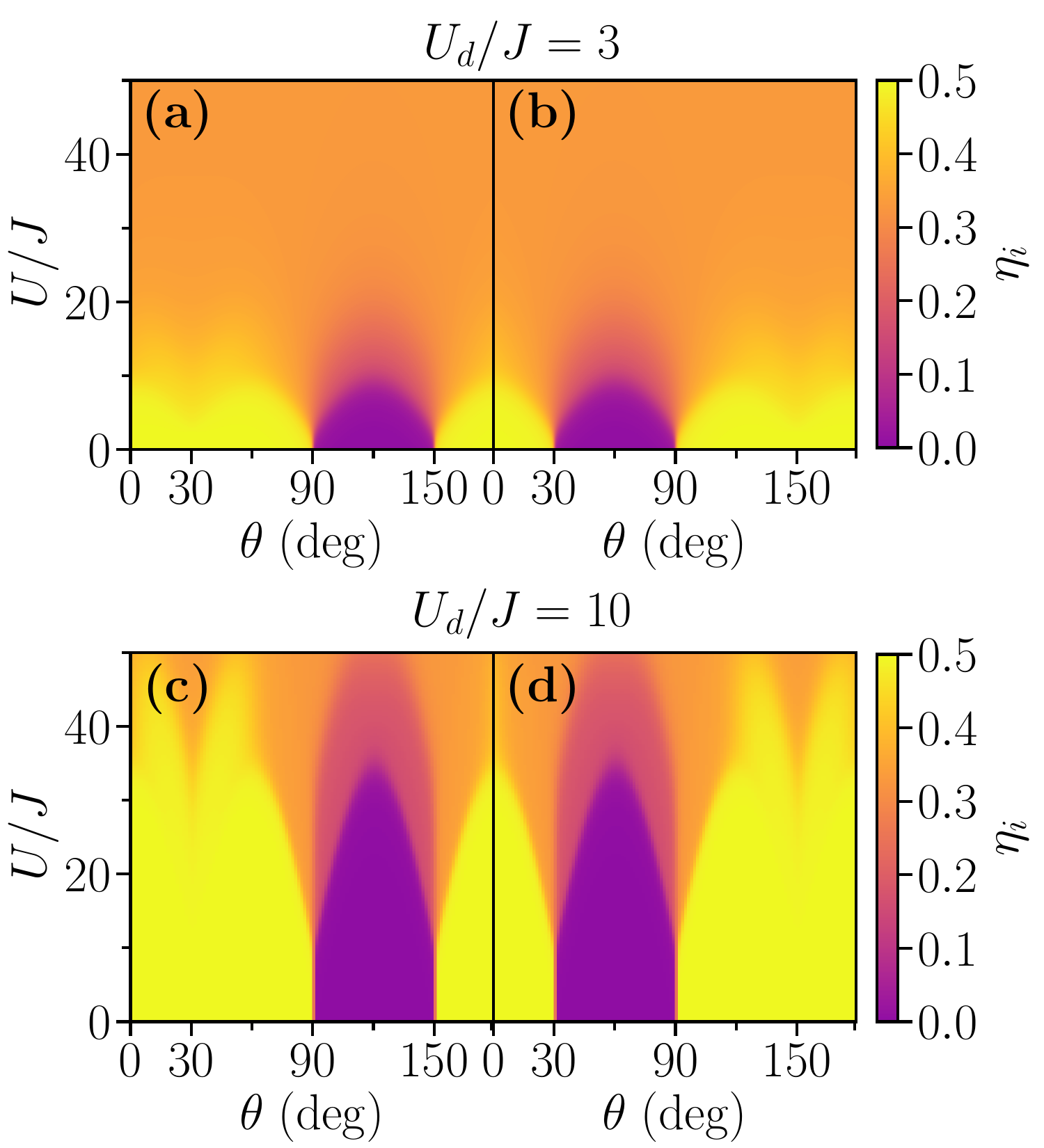}
    \caption{
    Average occupation  
    in the $(\theta, U/J)$ plane obtained from  diagonalization for $N=6$, $U_d/J=3$ (a, b), and $U_d/J=10$ (c, d). The color map shows $\eta_i$, the normalized average occupation in the ground state $\eta_i =  \bra{\Psi_{\rm GS}} \hat{n}_i \ket{\Psi_{\rm GS}}/N$,  of site 1 (a), (c) and site 2 (b), (d).}
    \label{occupationN6}
\end{figure}
The distribution of atoms between sites depends on the number of trapped atoms, and the relation 
between on-site (repulsive) interaction, tunneling, and inter-site anisotropic dipolar interaction. The sign and strength of the latter depends on the polarization angle and the site, see 
Fig.~\ref{tripleWellSetup_Vdd} (b).

The normalized average occupation of the $i$th-site over the ground state, 
$\eta_i =  \bra{\Psi_{\rm GS}} \hat{n}_i \ket{\Psi_{\rm GS}}/N$, 
is shown in Fig.~\ref{occupationN6}
for sites 1 (a), (c) and 2 (b), (d), for $N=6$, as a function of the dipolar angle and the on-site interaction. 
Note that there is a symmetry under $60$ deg rotations, that is, site 2 behaves as site 1 when $\theta = \theta+60$.
Occupation in (a), (b) panels ((c), (d) panels) correspond to a system with $U_d/J=3 \, (10)$. 
The results for different dipolar strengths are qualitatively similar since the effect of larger dipolar interaction strength resizes the range of small, intermediate, and large on-site interactions; even though lowering the effect of the hopping terms make transitions more abrupt.

In the dipole-dominating regime 
($U_d/J \gg U/J \gg 1$), 
and within the range of dipole angles where 
the inter-site interactions are repulsive concerning a 
particular site,
the average occupation in that site tends to zero. Meanwhile, the dipole-attractive sites will evenly distribute the atoms.
This is shown in Fig.~\ref{occupationN6}
for an even number of particles $N=6$ and $\nu=2$. 
When
$0< \theta < 30$ and $150 < \theta < 180$, the interaction between sites 1 and 2 is attractive, resulting in a ground state $\ket{\Psi_{\rm GS}}=\ket{N/2,N/2,0}$ (see the yellow-light regions with $\eta_i\simeq 0.5$ in both, left and right, panels).
Between
$30 < \theta < 90$,
the preferred sites are 1 and 3, and $\ket{\Psi_{\rm GS}}=\ket{N/2,0,N/2}$ (dark region in the right panels denoting that site 2 is empty). Between 
$90< \theta < 150$, the attractive sites
are 2 and 3, and
$\ket{\Psi_{\rm GS}}=\ket{0,N/2,N/2}$ (see the dark region in left panels of Fig.~\ref{occupationN6}).

In the opposing limit scenario, where the on-site interaction ($U/J \gg U_d/J \gg 1$) dominates the system, since it is repulsive, the three sites are equally populated, resulting in a Mott-like ground state: $\ket{\Psi_{\rm M}} = \ket{N/3,N/3,N/3}$, with $\eta_i=1/3$ for all sites.

\begin{figure}[hbt]
    \centering
    \includegraphics
    [width=246pt]{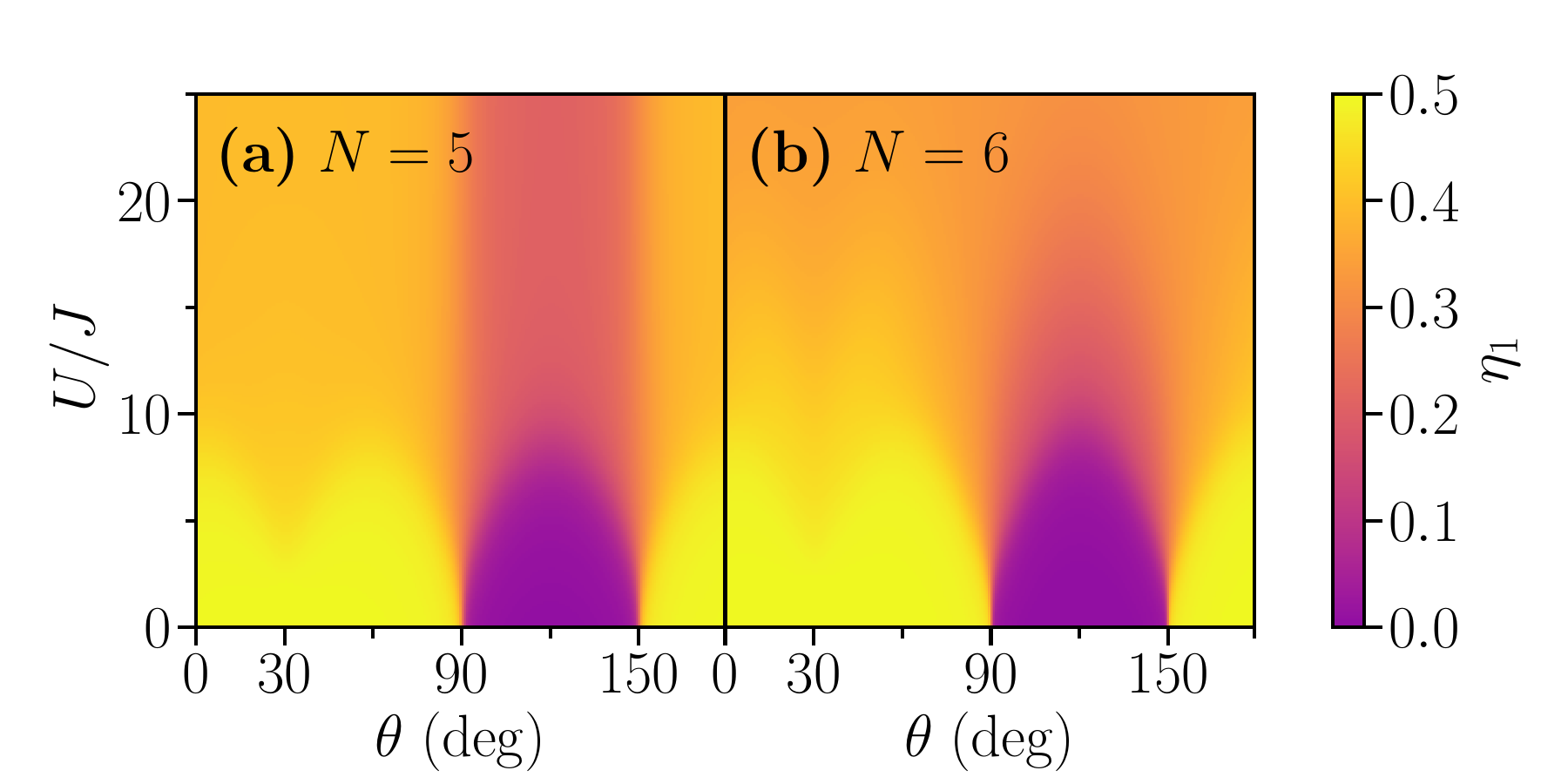}
    \caption{Normalized average occupation of site 1, $\eta_1=\expval{n_1}/N$, as a function of $U/J$ and dipole direction, with $U_d/J = 3$ for (a) $N=5$, and (b) $N=6$.}
    \label{ExpParticleNumSite}
\end{figure}

Figure \ref{ExpParticleNumSite} depicts the normalized average occupation of site 1 for $N=5$ (a) and $N=6$ (b).
For an odd value of $N$, within the dipole dominating regime, and across the range of dipole angles where 
the inter-site interactions is repulsive with respect to one particular site,
the average occupation in the latter site tends to zero while the remaining sites host the particles as equally as possible, with a normalized average occupancy of $1/2$. 
This is reflected in Fig.~\ref{ExpParticleNumSite}
with the purple-colored region between $90$ and $150$ deg, representing the empty site, while the yellow-colored regions of the remaining angles represent the average occupancy of 1/2. 
The wave function
in the dipolar-dominated regime is not a single Fock state, but a combination. For example, for 
$N=5$ and $\theta=0$, the ground state wave function is $\ket{\Psi_{\rm GS}}=(\ket{3,2,0}+\ket{2,3,0})/\sqrt{2}$.
The extra particle, with respect to the even case, is delocalized and is distributed amongst the dipole-favored sites, resulting in an entangled state~\cite{Escriva2019}.

On the limit of large on-site interaction ($U/J \gg U_{d}/J$), the particles tend to distribute equally between all sites to minimize contact interaction. 
Two cases arise: (i) Non-commensurate number of particles (fractional filling) where the favored sites have larger occupation values than the repulsive one as seen in Fig.~\ref{ExpParticleNumSite} (a). Note that the imbalance will remain for 
$U/U_d \gg 1$ 
as long as $U_d\neq 0$, 
this appears as an elongated dark shadow; 
(ii) Commensurate number of particles (integer filling) where all sites have an average occupation value of $1/3$ as seen in Figs.~\ref{occupationN6} and \ref{ExpParticleNumSite} (b).

At intermediate values of $U$ and larger values of $N$, there will be changes in the average occupation as the on-site interaction increases and the system minimizes the particle pairs until it reaches its minimum energy state. For example, for an incommensurate particle number such as $N=8$, as $U$ increases, different ground states are obtained at $\theta=0\,$: $\ket{4,4,0} \rightarrow \ket{4,3,1}+\ket{3,4,1} \rightarrow \ket{3,3,2}$. At every transition the average occupancy decreases until it reaches the Mott-like state. Interestingly, these results show that by manipulating the dipole orientation one can modify, at will, the population distribution of the system in the dipole-dominating range.

\subsection{Atomic limit in commensurate systems}

We consider the atomic limit by setting the tunneling to zero in a commensurate system. When $J=0$, the on-site and dipole interaction terms lie on the diagonal of the Hamiltonian. This allows us to obtain an analytical expression for the critical $U$ between the Mott-like state and the dipole-favored regime as a function of the dipole angle and the dipole strength.
The transition can be determined
by comparing the energy of the Mott-like state
dominated by the on-site interaction,
$\ket{N/3, N/3, N/3}$,
and the energy of dipole-favored states, for instance sites 1 and 2,
$\ket{N-n_i, n_i, 0}$, with $i=2$. The energy of these states can be calculated in the atomic limit as:
 \begin{eqnarray*}
     E_{\ket{N-n_i, n_i, 0}} &=& \frac{U}{2}[(N-n_i)^2 +n_i^2 -N] \\
     &+& 
     %C_{dd}
     U_d [1-3\cos^2(\theta_{12})](Nn_i-n_i^2) \,,
 \end{eqnarray*}
 \begin{eqnarray*}
     E_{\ket{N/3,N/3,N/3}} &=& 3\frac{U}{2} [N/3\left(N/3-1\right)] \\
     &+& U_d (N/3)^2 \sum_{i}^3\sum_{j>i}^3 (1-3\cos^2(\theta_{ij})) \\
     &=& \frac{N}{2} \left[U(N/3-1)-U_dN/3\right]\,.
 \end{eqnarray*}
 By imposing the two energies to be equal, it yields the critical contact interaction $U_c$ that determines the frontier between the two competing terms for a given polarization angle. The threshold that separates both regimes is:
 \begin{equation}
     U_c(\theta)     = \frac{U_{d}\left[N^2 + 6 (N - n_i)n_i (1-3\cos^2\theta)\right]}{2 \,[3(N-n_i)n_i - N^2]} \,,
     \label{Uc-1}
 \end{equation}
 where $\theta=\theta_{12}$ represents the dipolar angle associated to the dipole-favored sites 
 ($-30 \leq \theta \leq 30$). 
 When $N$ is even, or in the thermodynamic limit for odd $N$ (when 
 $N \rightarrow \infty$, and $n_i \rightarrow N/2$), the critical $U_c$ only depends on the dipole strength and orientation:
 \begin{equation}
U_c(\theta)
     = U_{d}\,(-5 + 9\cos ^2\theta) \,.
     \label{Uc-2}
 \end{equation}

\section{Entanglement properties}
\label{sec-entanglement}

At absolute zero and in absence of interactions, all the atoms of a bosonic gas populate the same single-particle state. Interactions
can remove particles from the single-particle ground state promoting them into excited states. This depletes a fraction of the condensate, or may even cause its fragmentation for sufficiently strong interactions. The condensed fraction of the system is represented by the largest eigenvalue of the one-body density matrix operator of the ground state:
\begin{equation}
    \hat{\rho}_{i,j} = \frac{1}{N}\bra{\Psi_{\rm GS}} \hat{a}_i^\dag \hat{a}_j \ket{\Psi_{\rm GS}}.
\end{equation} 
For a singly condensed system, the largest eigenvalue is $p_1 \approx 1$ while $p_2,p_3 \approx 0$, with $p_1+p_2+p_3 =1$. On the contrary, for a fragmented system, two or more eigenvalues will be comparable. The eigenvectors of the one-body density matrix are the so-called natural orbits or single-particle states.

Further insight on the quantum features 
and correlations can be obtained by analyzing entanglement properties. These attributes can be explored by computing the von Neumann entropy, which allows quantifying the correlation between sites. Moreover, it can also identify the configurations in which more than one Fock state is equally probable \cite{PhysRevA.87.053620,Gallem__2015}.

These correlations are obtained by calculating the reduced density matrix of one site by splitting the system in two parts and tracing out one of them. In our system, there are three possible bipartite splittings: (1,23), (2,13) or (3,12). 
Tracing out sites 2 and 3, results in the reduced density matrix of site 1, 
\begin{equation}
\rho_{1} =\Tr_{23}(\rho) =\sum_{n_1} \lambda_{n_1} \ket{n_1}\bra{n_1}, 
\end{equation}
where $\rho = \ket{\Psi_{GS}}\bra{\Psi_{GS}}$, and $\lambda_{n_1}$ are the Schmidt coefficients. The difference between the two largest Schmidt coefficients represents the Schmidt gap ($\Delta \lambda$), which is a useful
magnitude to detect phase transitions between different ground states \cite{gallemi2013,  Gallem__2015}.

With the reduced density matrix, the bipartite entanglement can be measured between the selected site $i$ and the rest of the system by computing the von Neumann entropy $ S_i = -\Tr(\rho_{i} \log \rho_{i})$.
Since the partial density matrix is diagonal, 
it can be rewritten as $S_i = -\sum \lambda_{n_i} \log \lambda_{n_i}$. If the subsystems exhibit entanglement, the entropy is maximized when at least two Schmidt coefficients are equal, resulting in a vanishing Schmidt gap $\Delta \lambda = 0$. On the contrary, if the subsystems are uncorrelated, only one non-zero Schmidt coefficient exists, resulting in zero entropy.

Due to the symmetry under rotation of the system, the correlations for subsystem (1,23) will be the same but shifted $60$ deg for the subsystem (3,12) and $-60$ deg for subsystem (2,13).

\subsection{Dipole strength effects}\label{dipoleStrength}
\begin{figure}[htb]
    \centering
    \includegraphics[width=0.98\columnwidth]{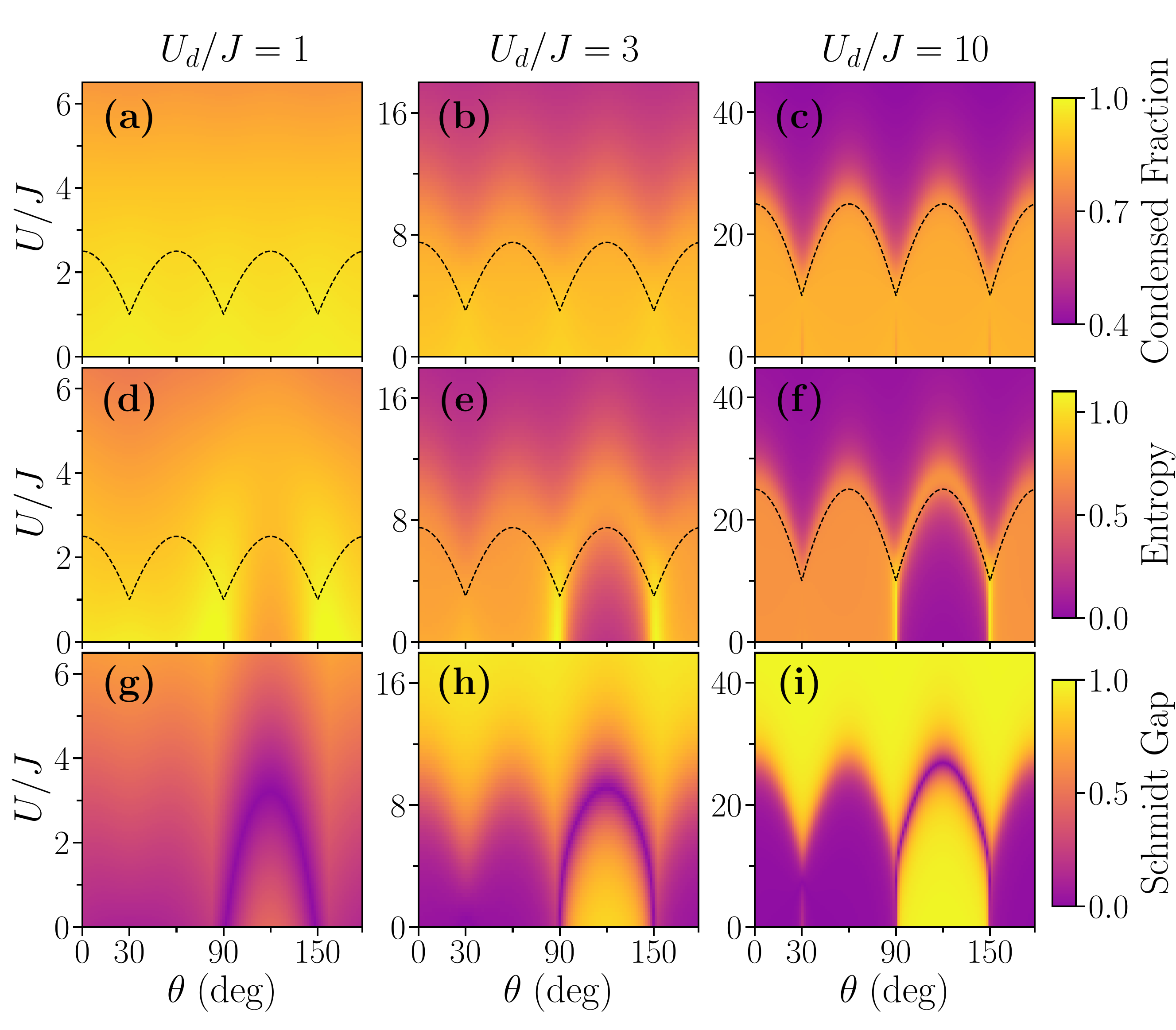}
    \caption{Results for $\nu = 1$ with each column representing different dipole strengths, from left to right $U_{d}/J=1,3$ and $10$. 
    (a-c): condensed fraction (largest eigenvalue of the one-body density matrix). (d-f): von Neumann entropy of $\rho_1$ (tracing sites 2 and 3).
    (g-i): Schmidt gap of $\rho_1$. All quantities are represented as color maps as a function of $U/J$ and the dipole direction.
    Dashed black lines correspond to the analytical values of $U_c$ where the ground state transition occurs in the zero-tunneling limit, Eq.~(\ref{Uc-1}). }
    \label{EntropySchmidtGap}
\end{figure}
Interesting effects arising from the dipole interaction already appear with $N=3$, filling factor $\nu=1$. Figure \ref{EntropySchmidtGap} shows the results on the condensed fraction, entropy, and Schmidt gap for different values of the dipolar interaction as a function of the polarization angle and $U/J$.
As the dipole strength increases (Fig. \ref{EntropySchmidtGap} panels from left to right), the required on-site repulsion strength ($U$) to fill the empty site, leading to an unfavorable dipolar interaction, increases, causing ground-state transitions to become more abrupt. 
The black-dashed line represents $U_c$ from Eq.~(\ref{Uc-1}), 
above which the condensed fraction decreases, highlighting the ground state transition, e.g., from a superposition of states $\ket{2,1,0} + \ket{1,2,0}$ to a Mott-like state $\ket{1,1,1}$. It should be noted that the eigenvectors change with the dipole angle. 

For small dipole strengths and in the noncontact interaction case ($U/J=0$), almost all particles are part of the condensate with $p_1 \approx 1$. As the on-site interaction increases there is a depletion of the condensate and at the large repulsion limit ($U/J\gg 1$) the system becomes fragmented with three single-particle states with the relative occupation values converging to $1/3$. For all dipole strengths and $U<U_c(\theta)$, there is a highly populated single-particle state,
with a large eigenvalue $p_1 \simeq 1$
showing that the dipole interaction might favor condensation. 

Regarding the correlation between sites, Fig.~\ref{EntropySchmidtGap} (d-i) shows the von Neumann entropy and the Schmidt gap tracing out sites 2-3. A strong correlation is obtained between the subsystems in the dipole-dominating regime demonstrated by the near-zero Schmidt gap and high entropy. Moreover, a sudden change in both values appears from 90 to 150 degrees. This can be understood with Figure \ref{occupationN6}, in which between those angles, site 1 is unfavorable and bosons only populate sites 2 and 3, breaking the correlation between the two subsystems of the partition. At the angles 90 and 150 which define the aforementioned frontier, a higher entropy value (lower Schmidt gap) appears, highlighting the relevance of this special configuration. The crossover between the dipolar dominant states that favors or disfavors the occupation of one specific site happens at $ \theta = 30$, $90$ or $150$ depending on the site, creating a highly correlated ground state for those configurations.
Therefore,
these specific orientations could be useful for applying protocols similar to the ones reported in Refs.~\cite{Wittmann2023,grun_protocol_2022} that aim to create entanglement in the dipolar dominant regime.

\subsection{Entanglement Spectrum}

\begin{figure*}[htb]
\centering
   \includegraphics[width=1\textwidth]{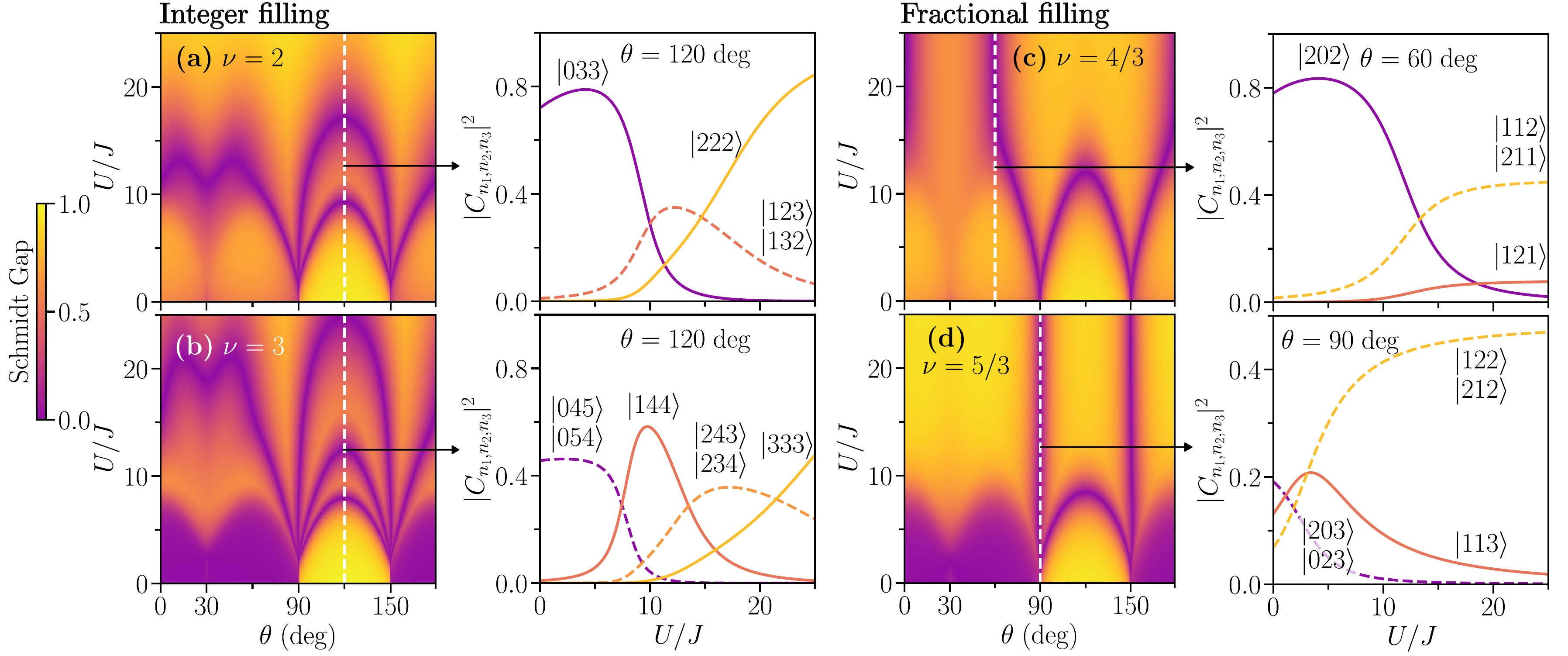}
    \caption{Schmidt gap of site 1 as a function of $U/J$ and the dipole angle, with a dipole strength of $U_{d}/J=3$ for different filling factors: For integer filling $\nu=2$ (a), $\nu=3$ (b) as indicated. For fractional filling $\nu=4/3$ (c), $\nu=5/3$ (d) as indicated. The white-dashed lines show the vertical cuts at which the ground state probabilities are
    depicted as a function of $U/J$ for a single dipole orientation: $\theta=120$ (a), (b); $\theta=60$ (c) and $\theta=90$ (d). Only relevant states are shown in the ground state probabilities as a function of $U/J$. The colored dashed lines represent the presence of two Fock states with identical weights in the ground state, highlighting the correlation between sites.}
    \label{IntegerFF}
\end{figure*}

To characterize the entanglement spectrum, we calculated the Schmidt gap corresponding to site 1 for different filling factors. The results are depicted in Fig.~\ref{IntegerFF}.

As a result of the anisotropy of the dipole interaction, the entanglement varies depending on the filling factor and whether the number of particles is odd or even. 
When the dipole interaction dominates
($U_d/J \gg U/J$)
the particles will be distributed among the dipole-attractive sites.
When $N$ is even, the latter are equally populated, as shown in the ground state probabilities,
$|C_{n_1,n_2,n_3}|^2$,
in Fig.~\ref{IntegerFF} (a) and (c), corresponding to $\nu=2$ and $4/3$\,, respectively. 
For $\nu=2$, the probabilities are calculated at $\theta=120$, where sites 2 and 3 attractive. Therefore, for large dipole interaction (weak on-site interaction) the most probable state is $\ket{0,3,3}$. Whereas for $\nu=4/3$ at $\theta=60$, the attractive sites are 1 and 3, and the most probable state in the dipole-dominated regime is 
$\ket{2,0,2}$.
These wavefunctions can be expressed as a product of states displaying no entanglement and a Schmidt gap close to 1 as shown in Fig.~\ref{IntegerFF}.

On the other hand, when $N$ is odd, within the region of large dipolar interaction, the particles cannot be equally distributed between the two dipole attractive sites. Then,
the ground state of the system will be a superposition of Fock states where the extra particle is shared between the dipole-favored sites, resulting in a wavefunction that cannot be expressed as a product of states. 
The superposition of states displays large correlations and entanglement between subsystems in the dipole angles where the studied site 
corresponds to a dipole-favored site.
This produces dark lobes in the Schmidt gap ($\Delta \lambda \simeq 0$) in Fig.~\ref{IntegerFF} (b), (d) ($\nu=3, \,5/3$). 
These lobes appear in the dipole-dominated regime for all values of the dipolar angle, except for $90< \theta < 150$ where 
sites 2 and 3 are attractive by the dipolar interaction and site 1 is empty and decorrelated with the others.

Figure \ref{IntegerFF} displays a vertical cut (white dashed lines) at a fixed dipole orientation at which the corresponding Fock state probabilities, $|C_{n_1,n_2,n_3}|^2$, of the ground state wavefunction are depicted as a function of $U/J$ (right panel). When there is a crossing of the highest probability Fock states, the Schmidt gap drops to 0 showing a strong correlation. Moreover, the correlations between sites are also reflected in the appearance of different Fock states with the same weight in the ground state (dashed lines in the right panels), caused by the symmetry of the system. Two Fock states can also be equally probable due to crossings.

For fractional fillings, Fig.~\ref{IntegerFF} (c) and (d) show two dark vertical stripes for large on-site interactions along which the Schmidt gap vanishes. These lines appear due to the localization of the non-commensurable particle in the dipole-favored sites, which change with the polarization angle and create a transition. 
When the dipolar angle is fixed, the Schmidt gap vanishes in the configurations where the are two Fock states with identical weight in the ground state (dashed lines in the ground state probabilities $|C_{n_1,n_2,n_3}|^2$ as a function of $U/J$).
For $\nu=4/3$ ($\nu=5/3$) at polarization angle $\theta=60$ ($90$), these Fock states are $\ket{112}$ and $\ket{211}$ 
($\ket{122}$ and $\ket{212}$)
 when $U/J > 15$ (5).

Additionally, it is interesting to stress that fractional filling factors can exhibit entanglement between sites regardless of on-site interaction at certain dipole angles, see Fig.~\ref{IntegerFF} (c) and  (d). That is because the ground state at these angles is a superposition that cannot be expressed as a product of states.
This appears, for example, for $\nu=5/3$ and $\theta=90$ or $\theta=150$; and for $\nu = 4/3$ at $\theta=0$ within polarization angles in the contact-dominant regime.

\section{Energy spectrum}
\label{sect-energy}

\begin{figure*}[ht]
    \centering
\includegraphics[width=1\textwidth]{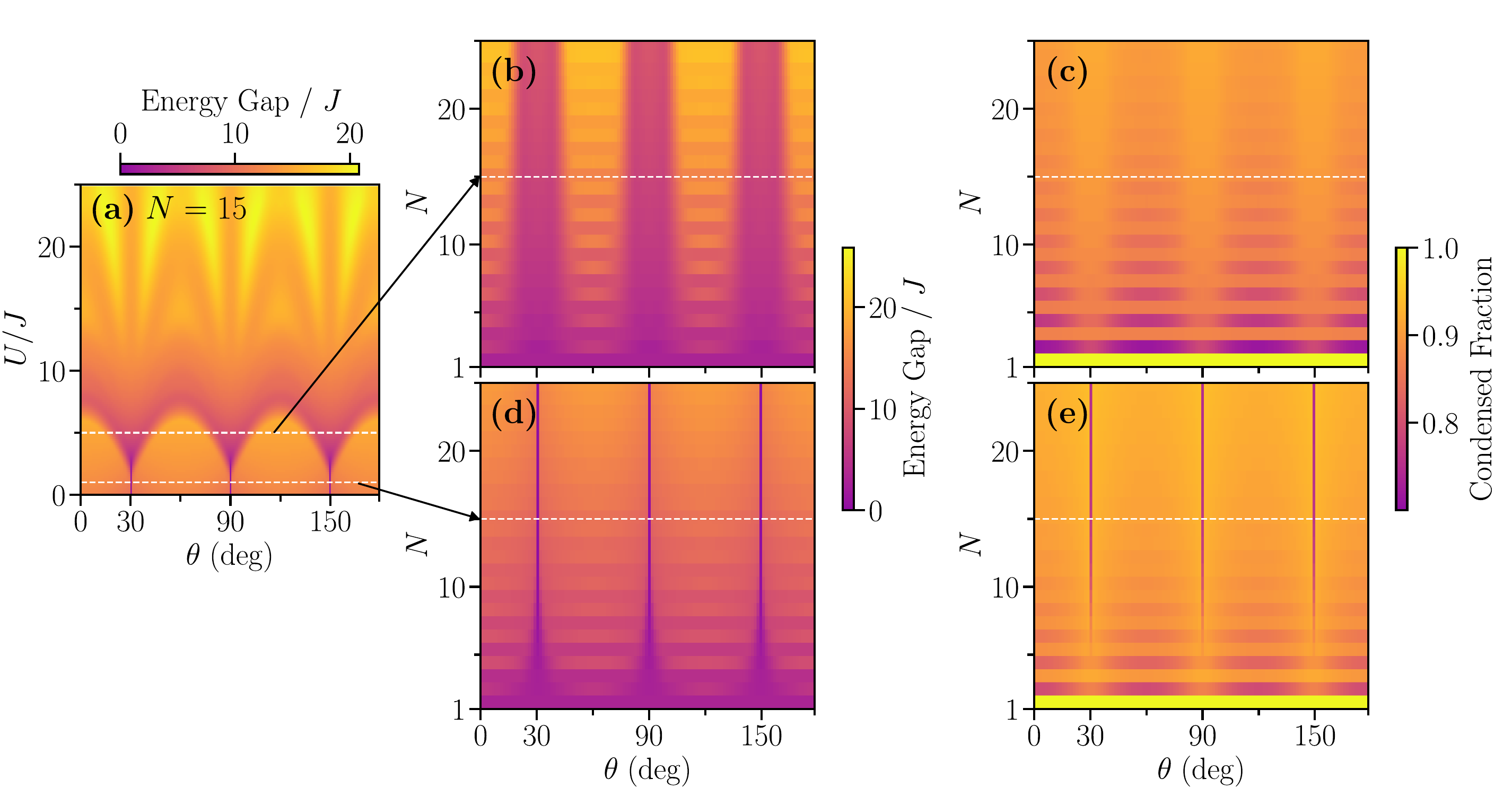}
\caption{(a) Energy gap as a function of $U/J$ and dipole angle with $N=15$ and dipole strength $U_{d}/J=3$. White dashed lines mark the values of $U/J$ used for (b), (c) ($U/J=5$) and (d), (e) ($U/J = 1$). Panels (b) and (c) are the energy gap and condensed fraction, respectively, as a function of the number of particles and the dipole angle, with a dipole strength $U_{d}/J=3$ and on-site strength $U/J=5$. 
Panels (d) and (e) are the energy gap and the condensed fraction, respectively, as a function of the number of particles and the dipole angle, with a dipole strength $U_{d}/J=3$ and on-site strength $U/J=1$.}
    \label{gapEnergy}
\end{figure*}

The degeneration in ground state energies can result in fragmentation of the system, as bosons can distribute in different single-particle states. In Ref.~\cite{Gallem__2015} it has been shown that a bi-fragmented state is achieved in contact-interacting bosons in a trimer when the tunneling between two sites becomes repulsive 
and the energy gap drops to zero. 
Since dipole interactions generate similar anisotropies between pairs of sites, see Fig.~\ref{tripleWellSetup_Vdd} (b), one can expect that bi-fragmented states can arise in our system in the dipole dominating regime.
In Fig.~\ref{gapEnergy} (a), we depict the energy gap between the ground state and the first excited state, as a function of $U/J$ and the dipole angle, for $N=15$
and $U_d/J=3$. In the region where the dipolar interaction dominates, there are certain values of the polarization angle ($\theta=30,\, 90,\, 150$) for which there is a zero gap energy, thus there are some special configurations where fragmentation could appear.

In Fig.~\ref{gapEnergy} (b) and (d), the energy gap evolution as a function of the number of particles shows a smooth behavior as the number of particles increases, highlighting that this degeneration is not a feature of the number of particles %and it 
but is due to the geometry and parameters of the system.

We show in Fig.~\ref{gapEnergy} (c) and (e) that although there is a zero energy gap, the system does not fragment, but a fraction gets depleted with the 
largest one-body eigenvalue $p_1 \simeq 0.76.$
This means that the energy gained by occupying one single-particle state is still sufficient to maintain a large population of the same single-particle state.
Nevertheless, Fig.~\ref{gapEnergy} (e) points out that at the thermodynamic limit within the highly dipolar-dominated region, it is possible that a bi-fragmented state arises at certain angles.

\section{Conclusions}\label{section:conclusions}

We have shown that triple-well potentials loaded with polar atoms with in-plane dipole orientation, present rich phenomena as a function of the polarization angle. Ground state properties strongly depend on the interaction strengths, number of particles, and dipole orientation. Our findings indicate that ground state crossing, occurring with varying dipolar interaction strength, is directly linked to correlations between sites. Moreover, we demonstrated that entanglement persists for incommensurate filling, regardless of the on-site interaction for certain dipole orientations.

We provide an analytic expression to calculate the on-site strength at which the system switches 
from a dipole-favored regime
into a Mott-like regime. It depends on the dipole interaction strength and orientation, and the number of particles. 
We found that, unlike on-site interaction, dipole interactions maintain  
the system with minimal depletion in the studied range of dipole strengths, which could provide an advantage in investigating superfluid currents. 

Moreover, we have shown that the dipole orientation can be used as a source for entanglement manipulation. This could be the ground for future investigations in the field of quantum technologies and a useful tool for developing entanglement manipulation protocols. Additional studies could go into exploring vortex currents within ring-shaped potentials, a phenomenon recently experimentally validated in a dipolar Bose-Einstein condensate~\cite{Klaus2022}. Moreover, extending the exploration of ground state and transport properties to larger systems of two triple-well rings with different tunneling connections and geometries could yield valuable insights. \\

\begin{acknowledgments}
This work has been funded by Grant No.~PID2020-114626GB-I00 from the MICIN/AEI/10.13039/501100011033,
by Grant No. 2021SGR01095
from Generalitat de Catalunya, and by Project CEX2019-000918-M
of ICCUB (Unidad de Excelencia María de Maeztu).
M. R. is supported by FPI Grant PRE2021-097235 and
H. B.-M. is supported by FPI Grant PRE2022-104397, both grants funded by MICIU/AEI/10.13039/501100011033 and by ESF+.
\end{acknowledgments}

\vspace{1pt}
\bibliography{references}% Produces the bibliography via BibTeX.

\end{document}